\documentstyle[aps,prb,epsfig]{revtex}
\begin{document}

\twocolumn[\hsize\textwidth\columnwidth\hsize\csname
@twocolumnfalse\endcsname

\title{Triplet superconductivity in quasi one-dimensional systems.}
\author{ A.A. Aligia$^{a}$ and Liliana Arrachea$^{b}$.}
\address{$^{a}$Centro At\'{o}mico Bariloche and Instituto Balseiro,\\
Comisi\'on Nacional de Energ\'{\i}a At\'{o}mica, \\
8400 Bariloche, Argentina,
$^{b}$ Departamento de F\'{\i}sica, 
FCyN Universidad de Buenos Aires\\
Pabell\'on I, Ciudad Universitaria, (1428) Buenos Aires, Argentina.\\
}
\maketitle

\begin{abstract}
We study a Hubbard hamiltonian, including a quite general
nearest-neighbor interaction, parametrized by repulsion $V$, exchange 
interactions $J_z, J_\perp$, bond-charge interaction $X$ and hopping 
of pairs $W$. The case of correlated hopping, in which
the hopping between nearest neighbors depends upon the occupation
of the two sites involved, is also described by the model for sufficiently
weak interactions.  
We study the model in  one dimension with usual continuum-limit field theory 
techniques, and determine the phase diagram. For arbitrary filling, 
we find a very simple necessary
condition for the existence of dominant triplet superconducting correlations 
at large distance in the spin SU(2) symmetric case: $4V+J<0$. In the 
correlated hopping model, the three-body interaction should be negative 
for positive $V$.
We also compare the predictions of
this weak-coupling treatment with numerical exact results for the 
correlated-hopping model obtained by
diagonalizing small chains, and using novel techniques to determine the 
opening of the spin gap.  
\end{abstract}


\vspace*{-1.4cm}
\vskip2pc]

\narrowtext

\section{Introduction}

In the last years, the study of extensions 
of the usual Hubbard model has, among others,  two important motivations.
First, to  explain many feaures of the rich diagram observed in the 
quasi-one dimensional organic compounds (TMTSF)$_2$X (where TMTSF 
means tetramethiltetraselenafulvalene and X represents, PF$_6$, ClO$_4$ or
other complex),   
other interactions must
be considered in the model hamiltonian, in addition to the 
on-site Coulomb repulsion \cite{bour,cam,voit,voitb}. 
Second, consensus increases about the
fact that the usual Hubbard hamiltonian
\cite{zhan}, does not seem to define
the minimal model for the superconducting cuprates, while  
additional terms, might explain the
physics of the superconducting phase 
of these materials \cite{dw}. 
The proximity between the spin density wave (SDW) and the 
superconducting phases
observed in the phase diagrams of both kind of systems has been
many times pointed out as a remarkable fact \cite{bour,cam}. 
The symmetry of the 
order parameter is, however, different for both systems. While
it has been observed to be $d_{x^2-y^2}$ for the case of the cuprates
\cite{arpes},
experimental evidence suggests that the observed superconductivity in
(TMTSF)$_2$ClO$_4$, and (TMTSF)$_2$PF$_6$ under pressure \cite{pw}, as well 
as in the layered compound Sr$_2$RuO$_4$  \cite{luke}
is of triplet $p$-wave character.
In this context, the extended Hubbard model with correlated hopping,
is particularly interesting, as it seems to provide a good scenario
for the occurence of both kind of superconducting instabilities, as well
as the SDW and charge density wave (CDW) ones, depending on the
values of the parameters and the filling factor \cite{dw,topo,prmon,japa}.
 
The model is defined by the hamiltonian
\begin{eqnarray}
&H& = \sum_{\langle i,j \rangle \sigma} ( c^{\dagger}_{i \sigma} c^{\;}_{j 
\sigma} + h.c. ) \left\{ t_{AA} (1-n_{i \bar{\sigma}}) (1-n_{j 
\bar{\sigma}}) \right. \nonumber \\ &&+ \left. t_{BB} \ n_{i \bar{\sigma}}n_{j 
\bar{\sigma}} + t_{AB} \left[ n_{i \bar{\sigma}} (1-n_{j \bar{\sigma}}) + 
n_{j \bar{\sigma}} (1-n_{i \bar{\sigma}}) \right] \right\} \nonumber \\ 
&&+ \ U \sum_i 
n_{i \uparrow} n_{i \downarrow} \;+\;
V \sum_{<ij>} n_{i }n_{j}  \ .
\label{1}
\end{eqnarray}

where $<ij>$ denotes nearest-neighbor sites. This model contains the one
which Hirsch proposed to give rise to hole superconductivity \cite{hir}.
In the context of the superconducting cuprates, 
Eq. (\ref{1}) has been obtained
as a one-band effective hamiltonian when a low energy reduction of the
extended three-band Hubbard hamiltonian is performed. 
For realistic parameters of the three-band model, 
calculated hopping parameters satisfy the following relation:   
$t_{AB}>(t_{AA}+t_{BB})/2$ \cite{fed,sim,pc1,physc}. 
In particular, in the limit
in which the Cu-O hopping $t_{pd}$ is much smaller than the charge-transfer
energies, $t_{AB}$ is of the order of $t_{pd}$ while $t_{AA}$ and $t_{BB}$ 
are of order  $t_{pd}^2$ \cite{pc1}.     
In two dimensions (2D), mean-field calculations, including the
effect of spin fluctuations,  
support the existence of a superconducting
phase with $d_{x^2-y^2}$ symmetry at finite doping, which is stabilized
with the addition of a next-nearest neighbor hopping in the hamiltonian
\cite{dw}.
While a realistic effective Hamiltonian for (TMTSF)$_2$X has not
been constructed so far, Eq. (1) contains the main terms, if only
the ground state with zero, one, and two particles in the singlet sector,
of an adequately chosen cell are retained \cite{sim,cell}. 

Collecting the correlated hopping terms can in one- two- and three-body
terms, the hamiltonian reads
\begin{eqnarray}
H &=&U\sum_{i}n_{i\uparrow }n_{i\downarrow }\;+\;
V \sum_{<ij>} n_{i }n_{j}  \nonumber \\
&+&\sum_{<ij>\sigma }(c_{i\bar{\sigma}}^{\dagger }c_{j\bar{\sigma}
}+{\rm H. c.}) [- t + t_2 (n_{i\sigma }+ n_{j\sigma })\nonumber \\
&+&
t_{3} \; n_{i\sigma} n_{j\sigma } ]\},
  \label{2}
\end{eqnarray} 
where $t=t_{AA},\; t_2=t_{AA}-t_{AB}$ and $t_3=2 t_{AB}-t_{AA}-t_{BB}$. 
While two-body interactions are usual in many-particle problems,
three-body interactions are more rare and introduce additional 
complications in most of the usual analytical many-body treatments.   
The appropriate mean-field reduction of the  
three-body term with parameter $t_3$ to effective two-body ones
can be performed using Wick's theorem and neglecting the resulting
normal ordered three body term, with the vacuum representing
the optimum Slater determinant \cite{pepe,blai}. 
In 1D, we have verified that in the continuum limit 
(representing the 
fermionic fields in terms of bosonic ones),
this procedure is equivalent to 
perform operator product expansions in the resulting hamiltonian, 
keeping only relevant and marginal operators \cite{japa}.  
In other words, both approaches are equivalent in the weak-coupling limit.
The ensuing two-body Hamiltonian reads \cite{physc}:
\begin{eqnarray}
H^{eff}& =& U\sum_{i}n_{i\uparrow }n_{i\downarrow }+
V^{eff} \sum_{<ij>} n_{i }n_{j}  \nonumber\\
&+& \sum_{<ij> \sigma }(c_{i\bar{\sigma}}^{\dagger }c_{j\bar{\sigma}}+{\rm H. c.})
[ t^{eff} + \Delta  (n_{i\sigma }+ n_{j\sigma })]\nonumber \\
&-&W \sum_{<ij> }(c_{i \uparrow}^{\dagger }
c_{i\downarrow}^{\dagger }c_{j\downarrow}c_{j \uparrow} + {\rm H. c.})
+ J \sum_{<ij>} {\bf S}_i \cdot {\bf S}_j,
\label{3}
\end{eqnarray} 
with
\begin{eqnarray} 
t^{eff}=t - t_3 (3 \tau^2 - \rho^2 ),\;\;\;
V^{eff}= V + t_3 \tau,\nonumber \\ 
W=2 t_3 \tau,\;\;\;  \Delta =t_2 + \rho t_3 ,\;\;\;  J=4 t_3 \tau,
\label {t3}
\end{eqnarray} 
where $\tau=\langle c_{i\sigma}^{\dagger }c_{j\sigma} \rangle$ and 
$\rho=\langle n_{i\sigma } \rangle=n/2$, with $n$ being the number
of particles per site. In 1D and the weak coupling limit:
\begin{eqnarray}
\tau &=& \frac{1}{\pi} \int_0^{k_F}dk  \cos k 
=\frac{\sin(\pi \rho)}{\pi},\nonumber\\
\rho&=& \frac{1}{\pi} \int_0^{k_F}dk= \frac{k_F}{\pi}.
\label{4}
\end{eqnarray}

The hamiltonian (\ref{3}), 
with arbitrary interactions $t^{eff}, \Delta , W, J,U, V^{eff}$,
defines the most general 
model with nearest neighbor two-body interations, 
which conserves charge and spin SU(2) symmetry. If in addition, 
the parameters are related by Eq. (\ref{t3}) with $V=0$, $\rho=1/2$ and
arbitrary $\tau$, the model also has pseudospin SU(2) symmetry \cite{afq}.
In the general case, we will also include the possibility of 
anisotropic exchange ($J_z$ in one direction, $J_\perp$ in the other two), 
breaking spin SU(2) symmetry.
Eq. (\ref{3}) contains all the contributions up to nearest-neighbors 
of the Coulomb interaction when written in the tight 
binding basis \cite{hub,voitl}. In this case, for weak screening 
of the interatomic repulsion, the relation
$U>V^{eff}>\Delta>W \sim J >0 $ has been derived, 
while smaller values of $V$ are expected if the screening is 
efficient\cite{hub,voitl,vol}. Arbitrary values of the 
different interactions could, however, be expected when
dealing with effective models, derived from some multiband   
model, as it is the case of Eq. (\ref{1}). 

Several specific cases of the model Eq. (\ref{3}) have been studied before 
using continuum limit field theory (CLFT) 
\cite{voitb,japa,voitl,frad,japa2}. In particular, 
the chain described
by the correlated hopping model Eq. (\ref{1}) was analized recently at
half filling \cite{japa}. However, no definite conclusions regarding possible 
dominance of superconducting correlations at large distances were obtained,
and the extension to other fillings remains open. An accurate phase diagram 
has also been obtained numerically using topological transitions \cite{topo}.
Other works on models similar to Eqs. (\ref{1}) and (\ref{2}) are cited in 
Refs. \cite{japa,fest}.
 
In this paper, we study the phase diagram of $H_{eff}$
(\ref{3}) with generic parameters 
($t^{eff}, t^{eff}_2, W, J_z, J_\perp, V^{eff}$), 
in the weak-coupling regime, using the CLFT. 
The boundaries of
the region with dominant triplet superconducting correlations
at large distances are given by simple analytical expressions. 
This region will be denoted TS phase in the following.
As a further step, we obtain the phase diagram of  the correlated hamiltonian
Eq. (\ref{1}), for weak and strong interactions, by numerical diagonalization
of finite rings. The opening of the spin gap is
detected accurately using a novel method based on results of conformal field 
theory and renormalization group \cite{naka}, which in turn is equivalent
to a topological transiton which corresponds to a jump in a Berry phase 
\cite{ali}. The results of both approaches are compared and the conditions
for the existence of the TS phase are discussed. 
We also analyze the other phases of the model, and discuss 
within which region of parameters, a
superconducting instability with $d_{x^2-y^2}$ symmetry could
be expected in 2D.  
Section II describes the weak coupling results.
Results of the exact diagonalization of Eq. (\ref{1}) and comparison
with the CLFT are presented in section III. Section IV contains a 
discussion.

\section{Continuum limit field theory.}
The CLFT, also called g-ology, is a weak coupling approach.
The whole procedure  has been explained in detail in many contributions
\cite{voit,voitb,japa,japa2,sol,wieg}. We, thus, present here only 
a brief explanation of the steps followed.
The basic  assumption is
that the interactions are small, in comparisson with the Fermi energy. 
The non-interating energy dispersion relation is linearized 
around the two Fermi points and the interactions in the momentum space
are expressed in terms of four different scattering processes, which
are labeled by coupling constants $g_{i||}$ ($g_{i \perp}$) if they involve 
the same (opposite) spins.

To describe the low-energy physics of the problem, the fermion operators
are decomposed as
\begin{equation}
 c_{j \sigma}  \rightarrow
\exp{(i k_F R_j)} \Psi_{ +, \sigma}(j) + 
\exp{(-i k_F R_j)} \Psi_{-,  \sigma}(j), 
\nonumber
\end{equation}
where $\Psi_{ r, \sigma}(j), \; r=^{+}_{-}$ describe
left- and right-moving fermions. 
In the continuum limit 
$a \rightarrow 0, \; L \rightarrow \infty $, with 
$ a L $ finite, being $a$ the lattice constant and $L$ the number
of sites of the lattice, these operators scale as
$\Psi_{ r, \sigma}(j) \rightarrow \sqrt {a} \Psi_{ r, \sigma}(x = j a)$.
The hamiltonian (\ref{3}) can be written in terms of the continuum
fields  $\Psi_{ r, \sigma}(x)$. 

 Using the same notation as Voit \cite{voit,voitb} and neglecting irrelevant
operators, we obtain the following coefficients for the different
scattering processes of the hamiltonian Eq. (\ref{3}):   
\begin{eqnarray}
g_{1 ||}&=& (2V^{eff} + \frac{J_z}{2}) \cos(2 k_F a), \nonumber\\
g_{2 ||} &=& g_{4 ||}= V^{eff}+\frac{J_z}{4}, \; g_{3 ||}=0 \nonumber \\
g_{1 \perp} &=& U+(2 V^{eff} - \frac{J_z}{2}) \cos(2 k_F a) 
-J_\perp \nonumber\\
& & -2W +  8 \Delta \cos( k_F a) \nonumber \\
g_{2 \perp} &=& \frac{U}{2}+V^{eff}-\frac{J_z}{4} 
- \frac{J_\perp}{2} \cos(2 k_F a)\nonumber\\
&-&W + 4\Delta \cos( k_F a), \nonumber \\
g_{3 \perp} &=& \delta_{n,1}(U-2V^{eff}+\frac{J_z}{2}+J_\perp+2W), \nonumber \\
g_{4 \perp} &=& \frac{U}{2}+V^{eff}-\frac{J_z}{4} 
- \frac{J_\perp}{2} 
-W\cos(2 k_F a)  \nonumber\\
&+&4\Delta \cos( k_F a), 
\label{g's}
\end{eqnarray}
where $\delta_{n,1}$ in the $g_{3 \perp}$ (Umklapp) coupling contant
indicate that this scattering process is active only at half-filling. 
The fermionic fields $\Psi_{ r, \sigma}(x)$ can be written in terms of 
bosonic fields $\phi_\rho$, $\phi_\sigma$, where $\rho,\; (\sigma)$ 
denotes charge (spin) degrees of freedom, by recourse to a bosonization 
identity \cite{voit} and the hamiltonian can be expresed as: 
\begin{equation}
H^{eff} = H_{\rho}+ H_{\sigma},
\end{equation}
where $H_{\nu}$,($\nu= \rho,\sigma$) is a sine-Gordon hamiltonian:
\begin{eqnarray}
H_{\nu} & =& v_\nu \int dx \{ \frac{1}{2}[\Pi^2_\nu(x)+(\partial_x \phi_\nu)^2]
\nonumber\\
& +&\frac{m_\nu}{a^2}\cos(\sqrt{8 \pi K_\nu} \phi_\nu)\},
\label{sg}
\end{eqnarray}
where $\Pi_\nu$ is the moment conjugate to $\phi_\nu$, and
\begin{eqnarray}
v_{\nu} = \sqrt{(v^{\nu}_F)^2-(\frac{g_{\nu}}{2 \pi})^2} , \;\;\;
m_{\rho} \;=\; \frac{g_{3 \perp}}{ 2 \pi}, \;\;\; 
m_{\sigma} \;=\; \frac{g_{1 \perp}}{2 \pi}, \nonumber\\
v^{\rho}_F \;=\; v_F + \frac{g_{4||}+g_{4 \perp}}{\pi} , \;\;\;
v^{\sigma}_F \;=\; v_F + \frac{g_{4||}-g_{4 \perp}}{\pi} \nonumber\\
v_F\;=\;2 t_{eff} \sin(\frac{\pi n}{2}), \;\;\;
K_{\nu}\;=\;\sqrt{\frac{2 \pi v^{\nu}_F + g_{\nu}}
{2 \pi v^{\nu}_F - g_{\nu}}},
\label{mass}
\end{eqnarray}
with
\begin{equation} 
g_{\rho} \;=\; 2g_{1 ||} - g_{2 ||} - g_{2 \perp}, \;\;\;
g_{\sigma} \;=\; 2g_{1 ||} - g_{2 ||} + g_{2 \perp}.
\label{g2}
\end{equation}
The physics of the sine-Gordon hamiltonian is well known from 
renormalization group \cite{voit,sol,wieg}.
At half-filling, for $|g_{3 \perp}| \leq g_{\rho}$ the charge sector 
renormalizes to the
Tomonaga-Luttinger fixed point, where charge excitations are gapless.
Away form half-filling, charge excitations are gapless, except, for 
commensurate fillings $n=p/q$ and strong repulsive interactions \cite{gia}:
the system is insulating if $K_\rho<1/q^2$. 
For example at quarter filling ($n=1/2$), the critical 
value of $K_\rho$ is 1/4 \cite{gia,gia2}.
In terms of the parameters of Eq. (\ref{3}), the condition for
Luttinger liquid behavior
at half filling reduces to:
\begin{equation}
2U+4V^{eff}+2J_\perp+J_z \leq 0 \; {\rm and}\; 2V^{eff} \leq W)  
\label{tlc}
\end{equation}
For the particular case of the extended Hubbard model with
correlated hopping (\ref{2}), using Eqs. (\ref{t3}),(\ref{4}) and (\ref{g's}),
this condition reads
\begin{equation}
(U+2V+8 t_3/\pi \leq \ 0 \; {\rm and}\; V\leq 0)
\label{tlcch}
\end{equation}
in agreement with Ref. \cite{japa}. In particular, for $V>0$ the system 
has always a charge gap at half filling.

Out of half filling, the hamiltonian $H_\rho$ (see Eqs. (\ref{sg}), 
(\ref{mass}), and (\ref{g's})), reduces to a gaussian model (except 
for higher order Umklapp processes which are relevant for commensurate
fillings and $K_\rho \leq 1/4$, as mentioned above \cite{gia,gia2}). 
In this case, $K_\rho$ is not renormalized and if $K_\rho > 1$
superconducting correlations dominate at large distances. 
From Eq. (\ref{mass}), this happens when  $g_\rho >0$. At half filling,
$K_\rho$ is renormalized, but the initial value (Eq. (\ref{mass})) should
be larger than one in order to reach a final value larger than one at 
the gaussian point in the renormalization-group procedure. Thus, from
Eqs. (\ref{4}), (\ref{mass}), (\ref{g's}) and (\ref{g2}), 
a necessary condition for the 
dominance of superconducting correlations at large distances is obtained: 
\begin{eqnarray}
& & -U-4 V^{eff}+(2 V^{eff} + \frac{J_z}{2}+J_\perp) \cos(\pi n) \nonumber \\
& & + 2W -  8 \Delta \cos(\pi n/2) > 0.
\label{cons}
\end{eqnarray}
This condition is also sufficient away from half filling ($n \neq 1$).
The character of the dominant superconducting correlations at 
large distances (singlet or triplet) and simultaneosly the opening of
a spin gap is determined by $H_\sigma$ (Eq. \ref{sg}) and the flow of 
its paramenters under renormalization. In the spin SU(2) invariant case
($J_z=J_\perp$), $g_\sigma=g_{1 \perp}$. For negative $g_\sigma$ a 
spin gap opens, the triplet superconducting (TS) correlations functions (CF)
decay exponentially, while the singlet superconducting (SS) CF decay as 
$d^{-1/K_\rho}$ with distance $d$. Instead, for positive $g_\sigma$, 
$K_\sigma$ renormalizes to 1, there is no spin gap, the TS CF decay as
$d^{-1-1/K_\rho}\ln^{1/2}d$ and dominate over the SS CF, which decay as
$d^{-1-1/K_\rho}\ln^{-3/2}d$ \cite{voit}.  Thus, in the Tomonaga-Luttinger
liquid phase with $J_z=J_\perp$ and $K_\rho >1$, 
using  Eqs. (\ref{4},\ref{g's},\ref{mass},\ref{g2}), one sees that 
TS CF dominate at large distances if and ony if:     
\begin{eqnarray}
& &U+(2 V^{eff} - \frac{J}{2}) \cos(\pi n) 
-J \nonumber \\
& & -2W +  8 \Delta \cos(\pi n/2) > 0.
\label{cont}
\end{eqnarray}
If the model is not spin SU(2) invariant, TS CF dominate, decaying as 
$d^{-1/K_\rho-1/K_\sigma}$ when $K_\rho>1$ and $g_\sigma>|g_{1 \perp}|$, which
in turn implies that the renormalized $K_\sigma>1$.

Adding Eqs. (\ref{cons}) and (\ref{cont}) a very simple necessary
condition for the existence of the TS phase in the model Eq. (\ref{3}) with
$J_z=J_\perp$ is obtained:
\begin{equation}
4 V^{eff}+J<0.
\label{consum}
\end{equation}
This result is consistent with theoretical analysis which relate triplet 
superconductivity with ferromagnetism \cite{sig}.

For the model with correlated hopping in the form of Eq. (\ref{2}), 
using Eqs. (\ref{t3},\ref{4}) the conditions (\ref{cons},\ref{cont}) 
take the form:
\begin{eqnarray}
& & -U+ 2 V (\cos(\pi n)-2)+ \frac{8}{\pi} t_3 \sin(\pi n/2)\cos(\pi n)
\nonumber\\
& & -8(t_2+t_3 n/2)\cos(\pi n /2) > 0,
\label{c1}
\end{eqnarray}
in order that superconducting CF dominate at large distances and:
\begin{eqnarray}
& & U+ 2 V \cos(\pi n) - \frac{8}{\pi} t_3 \sin(\pi n/2)\nonumber \\
& & +8(t_2+t_3 n/2)\cos(\pi n /2) > 0.
\label{c2}
\end{eqnarray}
for the region in which the spin gap is closed.
Adding both conditions leads to:
\begin{equation}
V + \frac{2}{\pi} t_3 \sin(\pi n/2) < 0,
\label{c3}
\end{equation}
as a necessary condition for the model to have a TS phase.
In the extended Hubbard model ($t_2=t_3=0$), Eqs. (\ref{c1}) and Eqs. 
(\ref{c2}) imply that a TS phase can only exist only for 
$2/3 \leq n \leq 4/3$.

In addition to the SS and TS CF, the phase diagram at half filling 
is determined by the  CF at large distances of the following order 
parameters for 
charge density wave (CDW), spin density wave (SDW), bond ordering wave
(BOW) and spin bond ordering wave (SBOW) order \cite{voit,japa}:
\begin{eqnarray}
O_{CDW} & = &\sum_{i \sigma} (-1)^i n_{i \sigma} \nonumber \\
 & \sim &
\cos(\sqrt{2 \pi K_\rho} \phi_\rho)
\cos(\sqrt{2 \pi K_\sigma} \phi_\sigma), \nonumber \\
O_{SDW} & = & \sum_{i \sigma} (-1)^i \sigma n_{i \sigma} \nonumber\\
& \sim &
\sin(\sqrt{2 \pi K_\rho} \phi_\rho)
\sin(\sqrt{2 \pi K_\sigma} \phi_\sigma), \nonumber \\
O_{BOW} & = & \sum_{i \sigma }
(c_{i+1 \bar{\sigma}}^{\dagger }c_{i\bar{\sigma}} + {\rm H. c.}) \nonumber \\
& \sim & 
\sin(\sqrt{2 \pi K_\rho} \phi_\rho)
\cos(\sqrt{2 \pi K_\sigma} \phi_\sigma), \nonumber \\
O_{SBOW} &= & \sum_{i \sigma } \sigma
(c_{i+1 \bar{\sigma}}^{\dagger }c_{i \bar{\sigma}} + {\rm H. c.}) \nonumber\\
& \sim &
\cos(\sqrt{2 \pi K_\rho} \phi_\rho)
\sin(\sqrt{2 \pi K_\sigma} \phi_\sigma),
\label{op}
\end{eqnarray}

The opening of the spin gap is accompanied by the ordering of  $\phi_s$ with 
expectation value  $\langle \phi_s \rangle=0$. When the charge gap is closed,
except for the particular case $g_\rho=|g_{3 \perp}|$ for which $K_\rho$
renormalices to 1, $g_\rho$ renormalices to a positive value and  $K_\rho >1$ 
at the fixed point. Then, as discussed above, TS (SS) CF dominate at large 
distances if the spin gap is closed (open). Instead, when the charge gap
opens ($|g_{3 \perp}| > g_\rho$) , 
$\phi_\rho$ orders with expectation value $\langle \phi_\rho \rangle=0$
($\langle \phi_\rho \rangle=\sqrt{\pi/(8 K_\rho)}$ ) for negative (positive)
$g_{3 \perp}$. If the spin gap is positive, this implies CDW (BOW) order, 
while if the spin gap is closed the dominant CF are are the SBOW (SDW) ones 
\cite{voit,voitb,japa}.
These considerations lead to the phase diagram shown in Fig. 1 for the
general model Eq. (\ref{3}) at half filling 
in the isotropic case $J_\perp=J_z=J$.
For $W=J=0$, the model reduces to the extended Hubbard model studied 
previously \cite{voitb,frad,hir2}, and no BOW or SBOW phases appear. For 
large positive $U$ the leading power-law decay of the SDW, CDW and BOW
CF is $1/d$, but the logarithmic correction of the former ($\ln^{1/2}d$), makes
the SDW the dominant CF at large distances \cite{voitb}. 
The combination of parameters which control the existence of the BOW and SBOW 
phases is $2W+J$. 
The BOW (BSDW) CF dominate in a certain region of $U$ and $V$ if
and only if $2W+J >0$ ($2W+J<0$), particularly for large $U$ near the line
$V=U/2-W-J/4$ at which the spin gap opens.
\begin{figure}
\narrowtext
\epsfxsize=3.5truein
\vbox{\hskip 0.05truein \epsffile{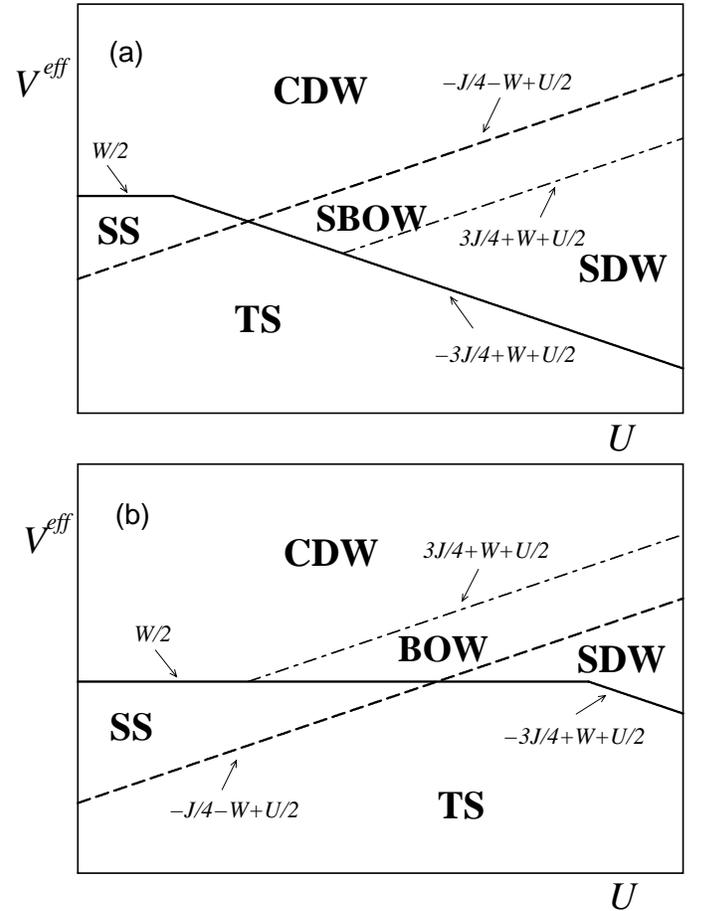}}
\medskip
\caption{
Phase diagram of the general model Eq.(\ref{3}) at half filling ($n=1$)
for $J_\perp=J_z=J$. (a) $2W+J<0$, (b) $2W+J>0$.}
\label{fig1}
\end{figure}

\section{Numerical results}
In this section we calculate the phase diagram of the correlated hopping 
model Eq. (\ref{1}) by exact diagonalization of finite rings, and compare 
the results with those of the previous section. We concentrate on the
electron-hole symmetric case and set $t_{AA}=t_{BB}=1$. 
Numerical \cite{topo} and CLFT \cite{japa} results for 
one particle per site ($n=1$) have been done recently. 
We have concentrated on other two  densities: 
$n=1/2$ (quarter filling) and $n=2/3$.
The boundary of the superconducting phase was determined from the equation
$K_\rho =1$, with $K_\rho$ calculated from the expression:
\begin{equation}
K_\rho =\sqrt{\pi \kappa D_\rho}/2,
\label{kro}
\end{equation}
where the Drude weigth $D_\rho$ and the compressibility $\kappa$ were obtained
extrapolating to the thermodynamic limit numerical results (obtained in the
usual way \cite{5})  using a polynomial in $1/L$, where $L$ is the length 
of the ring. For $n=1/2$, we have used rings with $L=8$, 12, and 16, and for 
$n=2/3$ the lengths used were $L=6$ and 12. We have also calculated $K_\rho$
using other two expressions which involve $v_\rho$, the central charge and 
charge and spin gaps \cite{5}, to check for consistency and finite-size 
effects. The latter are in general very small, except for specific cases 
mentioned below.
 The opening of the spin gap was determined from the crossing of the  
levels of lowest energy in the sectors with total spin $S=0$ and $S=1$ for
periodic (antiperiodic) boundary conditions if $N/2$ is even (odd), where
$N$ is the number of particles in the system \cite{naka}. This method is 
based on results of conformal field theory and renormalization group,
which show in addition that at the crossing point, the finite size 
corrections of these excitation energies per site go as $1/L^2$, without
logarithmic corrections. This allows a very accurate determination of the
parameters for which the spin gap opens using finite-size scaling. This
level crossing has also a topological significance, since at this point, the
spin Berry phase, which can only take two values: 0 or $\pi$ (mod $2\pi$) 
jumps in ($\pi$) at the transition \cite{ali}.

 In Fig. 2, we show the evolution with doping of the phase diagram for $V=0$.  
As soon as the density decreases below $n=1$, the Umklapp processes become 
irrelevant and according to the CLFT, the degeneracy of TS and SBOW CF on one
phase and the SS and CDW on the other \cite{topo,japa} is broken in favor
of the supercondunting CF, since $K_\rho$ becomes larger than one. As doping
increases, the SDW phase advances over the TS phase and at quarter filling 
($n=1/2$) the TS phase lies entirely within the region of negative $U$. 
Instead the SS phase advances rapidly over the BOW (or dimer ordered) 
phase at half filling. The agreement of the predictions of the singlet-triplet
level crossing method for detecting the opening of the spin gap with the
CLFT results is excellent at weak coupling ($|t_{AB}-1|<0.1$), what
confirms the accuracy of the method. 
The numerical results for the points
at which $K_\rho=1$ also agree very well with the weak coupling results for
$n=1/2$. For $n=2/3$, for which only two points ($L=6$ and $L=12$) 
were used in the finite-size scaling, and $K_\rho$ varies very slowly with
the parameters near the non-interacting limit, the numerical points in the
weak coupling limit are not accurate enough. However, these points are also
consistent with the CLFT results. 
\begin{figure}
\narrowtext
\epsfxsize=3.5truein
\vbox{\hskip 0.05truein \epsffile{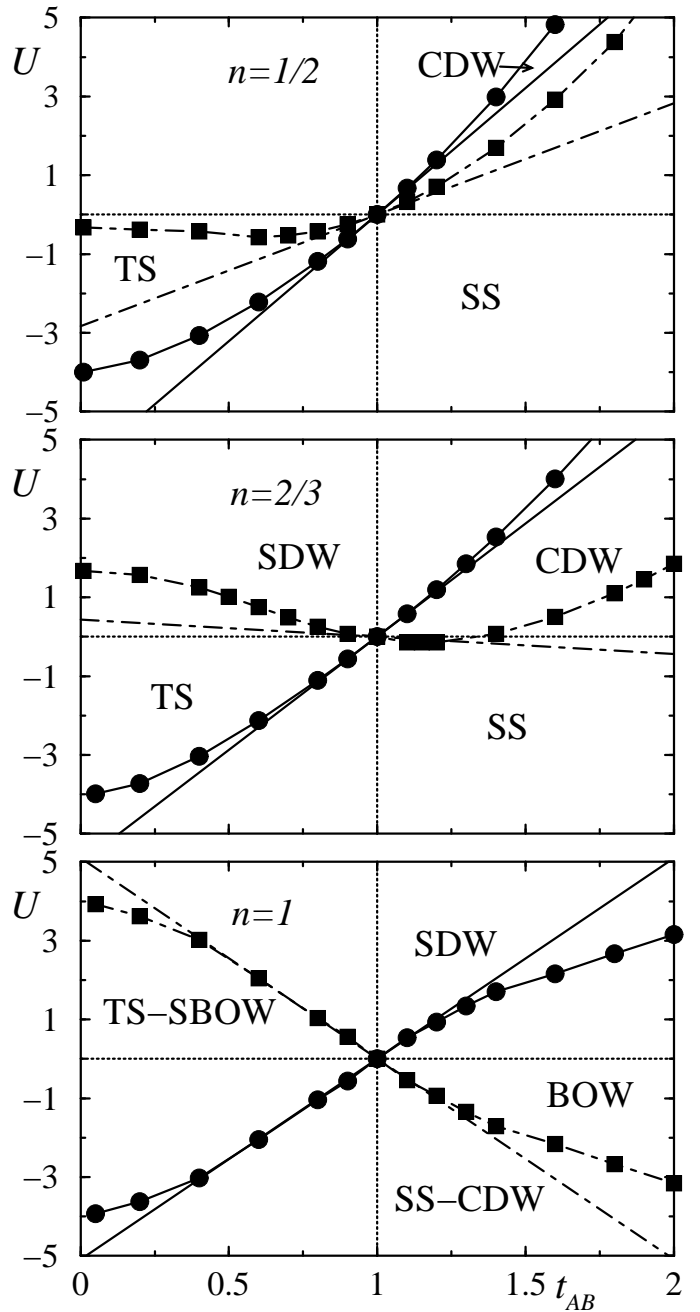}}
\medskip
\caption{Phase diagram of the correlated hopping model Eq.(\ref{1}) for $V=0$ and
several densities indicated at the top left of each figure. Solid squares 
indicate values of $U$ and $t_{AB}$ for which $K_\rho=1$. The straight line 
$K_\rho=1$ according to the CLFT 
(Eq. (\ref{c1}) with $t_2=1-t_{AB}$, $t_3=-2t_2$) is shown dot dashed. 
Solid circles
are points at which the spin gap closes. The corresponding results according
to the CLFT (Eq. (\ref{c2})) are represented by the full straight line.}
\label{fig2}
\end{figure}

It is remarkable that for $n=1$, the CLFT results are quantitatively valid 
even at intermediate coupling. 
However, out of half filling and weak coupling, the
CLFT overestimates the region in which the spin gap vanishes, and clearly
underestimate the extension of the superconducting phases.

\begin{figure}
\narrowtext
\epsfxsize=3.5truein
\vbox{\hskip 0.05truein \epsffile{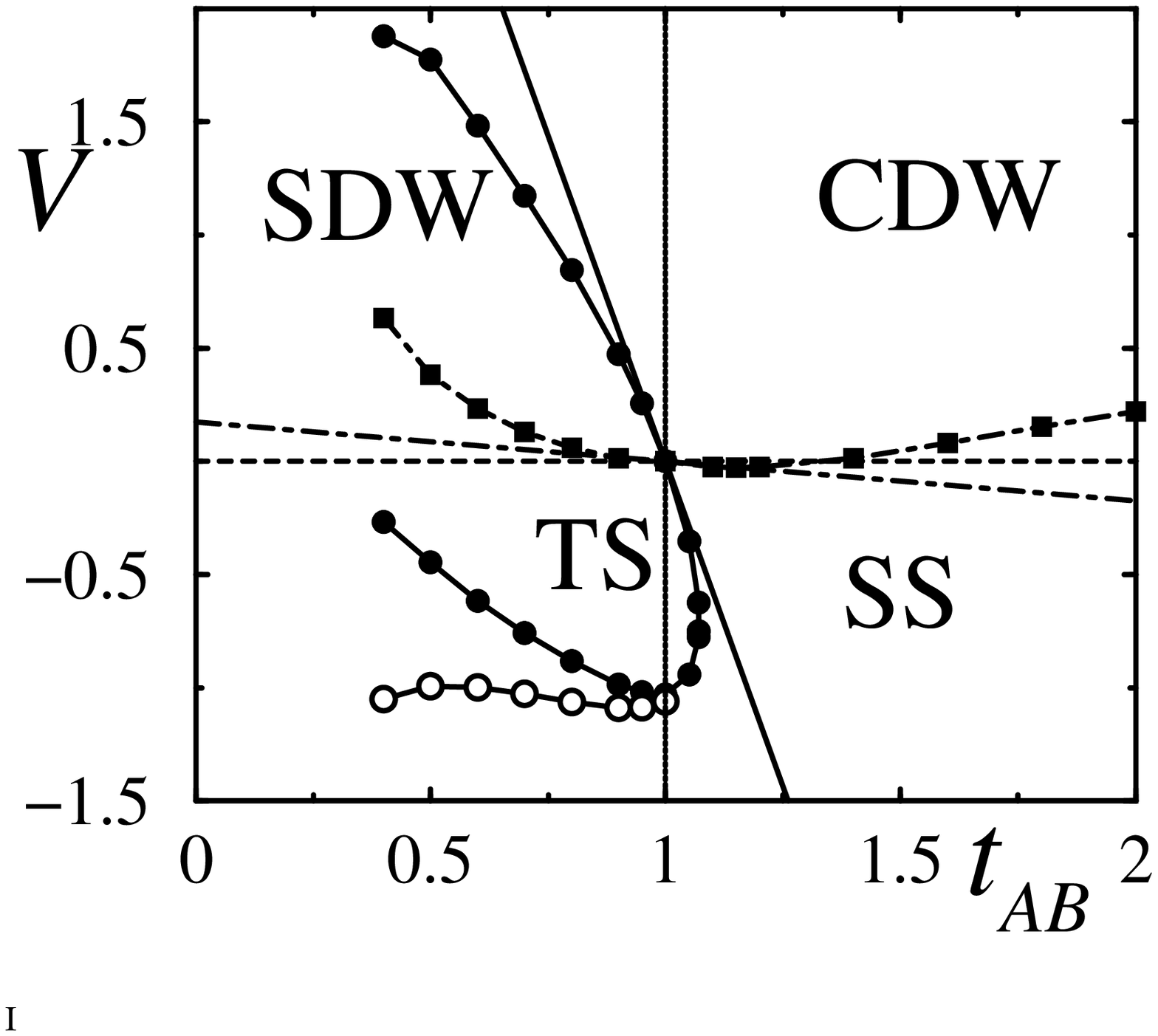}}
\medskip
\vspace*{-0.85cm}
\caption{Same as Fig. 2 for $U=0$ and $n=2/3$. The open circles are points which
correspond to the singlet-triplet crossing for $L=12$ and deviate 
substantially from the
extrapolated results that indicate the opening of a spin gap.}
\label{fig3}
\end{figure}

In Fig. 3 we show the effect of $V$ on the phase diagram for $n=2/3$. We 
have chosen this density because a TS phase exists for positive $U$ if 
$t_{AB}<1$. For $t_{AB}<0.4$ the results are affected by large finite size 
effects, perhaps because of the proximity of phase separation (PS), 
and we were 
unable to obtain reliable results. For $0.4 \leq t_{AB} <1$, increasing $V$,
from zero, the line of $K_\rho=1$ is crossed first and then at larger $V$ a
spin gap opens, in qualitative agreement with the CLFT results. However, we
obtain that the spin gap opens also when $V$ is decreased taking negative
values. This crossing is out of the reach of the CLFT. As $V$ is further
decreased one expects PS. For $t_{AB}<0.9$ our results 
for the opening of the spin gap at negative $V$ are affected by large finite
size effects (reflected by the difference between the values of $V$ 
indicated with solid and open circles in Fig. 3). This is probably caused
by the proximity to PS. The numerical investigation of PS is very delicate
\cite{clay} and is beyond the scope of this work.

\section{Discussion}
We have studied the phase diagram of a Hubbard model Eq. (\ref{3})
in the weak coupling limit, generalizing previous studies
which use the continuum limit field theory. The model 
includs the
most general form of nearest-neighbor two-body interactions which conserves
charge and spin. The phase diagram is very rich, and six different
phases can appear, according to the dominant correlation functions at 
large distances (see Fig. 1). In particular, in the isotropic case, if there
is a nearest-neighbor attraction ($V<0$) or ferromagnetic exchange ($J<0$) a
phase with dominant triplet superconducting correlations (TS phase) can
exist. 
Specifically $4V+J<0$ is a necessary condition for the existence of the 
TS phase. Work on weakly coupled chains using perturbation theory suggests that
small interchain hopping stabilizes a 3D long-range order with finite 
critical temperature, which corresponds to the dominant correlations at 
large distances in the purely 1D case \cite{fir}. Thus, we expect that 
our results can be applied to real quasi one-dimensional materials, 
which can be described by an effective Hamiltonian like Eq. (\ref{1}) or 
(\ref{3}).  
Since in real materials $V$ is expected to be repulsive, 
an efficient screening of the interatomic repulsions and effective 
ferromagnetic exchange (like that present in one dimensional cuprates
containing edge-sharing CuO$_4$ units \cite{mizu}) would be necessary 
conditions for the existence of triplet superconductivity. 
It has been proposed that in some ideal 
limit, ferromagnetism and triplet superconductivity might be related by
symmetry operations of the group SO(5) \cite{sig}.

For weak coupling, the general model Eq. (\ref{3}), with parameters 
satisfying the relations Eq. (\ref{t3}) describes the correlated hopping model
Eq. (\ref{1}). For this model to display a TS phase in the weak 
coupling limit, 
it is necessary that 
$V + (2/\pi) (2 t_{AB}-t_{AA}-t_{BB}) \sin(\pi n/2) < 0$. We have
studied this model beyond the weak coupling regime by numerical 
diagonalization of finite rings. In spite of the fact that the size of the 
studied systems is small, the results agree very well with those obtained
with the field theory in the weak coupling regime. In particular, the 
accuracy of determining the gap by the method of the crossing of singlet
and triplet excitations \cite{naka} (which in turn is equivalent to a 
topological transition in the spin Berry phase \cite{ali}) is confirmed.
For values of the correlated hopping which are outside the reach of the 
field theory, the regions of dominating superconducting correlations at
large distances extend beyond the predictions of the  weak coupling
treatment.

There are other physical phenomena which are outside the scope of 
the field theoretical treatment we followed. One of them is the opening
of the spin gap for $t_{AB} <1$ and negative $V$ found in our 
numerical calculations (see Fig. 3). Another one is phase separation.
In addition, for very small $t_{AB}$ ($t_{AB}=0.2$), there is numerical
evidence of peaks at incommensurate wave vectors in charge-charge and 
spin-spin correlation functions at half filling \cite{prmon}. 
These can be qualitatively understood using the formalism of the exact
solution for $t_{AB}=0$ \cite{afq}: roughly, the doubly occupied sites 
are represented by effective bosons and the singly occupied sites by
effective fermions. The Fermi wave vector of these effective fermions 
depends on $U$ and hence, it is in general different from the non-interacting 
Fermi wave vector. Unfortunately, a quantitative analytical calculation of 
these correlations functions for small $t_{AB} \neq 0$, is very difficult due 
to the huge degeneracy for $t_{AB}=0$. On the level of the field theory, one
might speculate that irrelevant operators, which we have neglected, become
important at large couplings in an adequate renormalization group treatment
and lead to incommensurations. 
In addition to the TS phase, the region of singlet superconductivity (SS) for
$t_{AB}>1$ and $U>0$  (see Fig. 2) is particularly interesting. In the case 
$t_3=2 t_{AB}-t_{AA}-t_{BB}=0$, SS has been proposed and found in a mean-field
treatment by Hirsch as a model for hole superconductivity \cite{hir}, and
confirmed by other numerical and analytical calculations 
\cite{japa,japa2,fest,5}. 
However, the case we studied numerically 
in section III is electron-hole symmetric and $t_3$
plays an important role. It is also essential in 2D to 
give rise to $d_{x^2-y^2}$-wave
superconductivity \cite{dw}. In the limit in which $t_{AB}$ is much larger 
than all
other energy scales of the system, at half filling, 
a reasonable approximation to the 
ground state is obtained splitting the ring into consecutive dimers, and
solving the Hamiltonian in each dimer. Including the hopping between 
dimers in second-order perturbation theory leads an energy which is above the
energy calculated with density-matrix renormalization group by $1.6\%$ 
\cite{topo}. This suggests a picture in which the system is composed of dimers,
which behave as hard core bosons, being frozen at half filling (leading to
a dimerized phase with long range order, BOW in Fig. (2)) but which aquire
mobility out of half filling, giving rise to dominant SS correlations. 
If this image can be extended
to 2D, we expect some kind of short range resonance-valence bond ground state 
at half filling, since dimers can be ordered in many different ways, and SS
of $s$- or  $d_{x^2-y^2}$-wave symmetry would naturally arise 
as the system is doped. The favored
symmetry depends on the topology of the Fermi surface. According to
mean-field calculations and including spin fluctuations,  the  
$d_{x^2-y^2}$-wave symmetry is favored for moderate doping if a negative
next-nearest-neighbor hopping is included in the model \cite{dw}. We are
presently investigating this possibility by numerical diagonalization.

\section*{Acknowledgments}
One of us (A. A. A.) thanks Fabian Essler for useful discussions. We thank 
the Max-Plank Institut f\"ur Physik komplexer Systeme, where this work
was finished, for its hospitality.
L. A acknowledges support from CONICET. A. A. A. is partialy supported
by CONICET. This work was supported by PICT 03-00121-02153 of 
ANPCyT and PIP 4952/96 of CONICET.


\begin{references}
\bibitem{bour} Bourbonais, Science {\bf 281}, 1155 (1998).
\bibitem{cam} D.K. Campbell, T. A. DeGrand, and S. Mazumdar,
 Phys. Rev. Lett. {\bf 52} 1717 (1984).
\bibitem{voit} J. Voit, Rep. Prog. in Phys.  {\bf 58}, 977  (1995).
\bibitem{voitb} J. Voit, Phys. Rev. B {\bf 45}, 4027 (1992).
\bibitem{zhan}Shiwei Zhang, J. Carlson and J. E. Gubernatis,
Phys. Rev. Lett. {\bf 78} 4486 (1999). 
\bibitem{dw}L. Arrachea and A. Aligia, Phys. Rev. B 
\underline{59} 1333 (1999).
\bibitem{arpes}
A. G. Loeser, Z.X. Shen, D.S. Dessau, D.S. Marshall, C.H. Park, P.
Fournier and A. Kapitulnik, Science {\bf 273}, 325 (1996); H. Ding, T.
Tokoya, J.C. Campuzano, T. Takahashi, M. Randeira, M.R. Norman, A.
Kapitulnik, T. Mochika, K. Kadowaki, and J. Giapintzakis, Nature {\bf 382},
51 (1996); G. Blumberg, M. Kang. and C. Kendziora, Science {\bf 278} 1427
(1997).
\bibitem{pw} I.J. Lee, M.J. Naughton, G.M. Danner, and P.M. Chaikin, 
Phys. Rev. Lett. {\bf 78}, 
3555 (1997); S. Belin and K. Behnia, {\it ibid} {\bf 79}, 2125 (1997).
\bibitem{luke} G.M. Luke, Y. Fudamoto, K.M. Kojima, M.I. Larkin, J. Merrin,
B. Nachumi, Y.J. Uemura, Y. Maeno, Z.Q. Mao, Y. Mori, H. Nakamura, and
M. Sigrist, Nature {\bf 394}, 558 (1998).     
\bibitem{topo}  A.A. Aligia, K. Hallberg, C.D. Batista, and 
G. Ortiz, cond-mat /9903213.
\bibitem{prmon}L. Arrachea, E. R. Gagliano and A. Aligia, Phys. Rev. B 
{\bf 55} 1173 (1997).
\bibitem{japa}  J. Japaridze and A. Kampf, Phys. Rev. B {\bf 59}, 12822
(1999); references therein.
\bibitem{hir} J.E. Hirsch, Physica C 158, 326 (1989); H.Q. Lin and J.E. Hirsch,
Phys. Rev. B {\bf 52}, 16155 (1995). 
\bibitem{fed}H.B. Sch\"{u}ttler and A.J. Fedro Phys. Rev. B {\bf 45}, 7588
(1992).
\bibitem{sim} M.E. Simon, M. Bali\~{n}a and A.A. Aligia, Physica C {\bf 206},
 297 (1993); M.E. Simon, A.A. Aligia and E. Gagliano, Phys. Rev. B {\bf 56},
 5637 (1997); references therein.
\bibitem{pc1} L. Arrachea and A.A. Aligia,  Physica C {\bf 289} 70 (1997). 
\bibitem{physc} L. Arrachea and A.A. Aligia, Physica C {\bf 303} 141 (1998).
\bibitem{cell} J.H. Jefferson, H. Eskes, and L.F. Feiner, 
Phys. Rev. B {\bf 45}, 7959 (1992).
\bibitem{pepe} J. Lorenzana, M.D. Grynberg, L. Yu, K. Yonemitsu, and A.R. 
Bishop, Phys. Rev. B {\bf 47}, 13156 (1993)
\bibitem{blai}J.-P. Blaizot and G. Ripka, {\it Quantum Theory of 
Finite Systems} (MIT, Cambridge, Massachusetts, 1986).
\bibitem{afq} L. Arrachea, A.A. Aligia, and E.R. Gagliano, Phys. Rev. Lett.
{\bf 76} 4396 (1996); references therein.
\bibitem{hub} D.K. Campbell, J. Tinka Gammel, and E.Y. Loh, Jr., 
Phys. Rev. B {\bf 42}, 475 (1990); references therein; A. Painelli and A.
Girlando, {\it ibid} {\bf 39}, 2830 (1989); A. Fortunelli and A. Painelli,
Chem. Phys. Lett. {\bf 214}, 402 (1993).
\bibitem{voitl} J. Voit, Phys. Rev. Lett. {\bf 64} 323 (1990).
\bibitem{frad} J.W. Cannon and E. Fradkin,
Phys. Rev. B {\bf 41}, 9435 (1990). 
\bibitem{vol} R. Strack and D. Volhardt, Phys. Rev. Lett.
{\bf 70} 2673 (1993).
\bibitem{japa2} G. Japaridze and E. M\"uller Hartmann, Ann. Phys. 
(Leipzig)  {\bf 3}, 163  (1994).
\bibitem{fest} A.A.Aligia, E. Gagliano, L. Arrachea, and K. Hallberg,
Eur. Phys. J B {\bf 5}, 371 (1998); references therein.
\bibitem{naka}M. Nakamura, K. Nomura, and A. Kitazawa, Phys. Rev. Lett.
{\bf 79}, 3214 (1997);M. Nakamura, J. Phys. Soc. Jpn {\bf 67}, 717 (1998). 
\bibitem{ali}A.A. Aligia, Europhys. Lett. {\bf 45}, 411 (1999).
\bibitem{sol} J. S\'olyom, Adv. Phys. {\bf 28}, 201 (1979)
\bibitem{wieg} P. Wiegmann, J. Phys. C {\bf 11} 1583 (1978);
D. Boyznovsky, J. Phys. A {\bf 22}, 2601 (1989).
\bibitem{gia} T. Giamarchi, Physica B {\bf 230-232}, 975 (1997).
\bibitem{gia2} T. Giamarchi and A.J. Millis, 
Phys. Rev. B {\bf 46}, 9325 (1992).
\bibitem{sig} S. Murakami, N. Nagaosa and M. Sigrist, cond-mat/9811001
\bibitem{hir2} J.E. Hirsch, Phys. Rev. Lett. {\bf 53}, 2327 (1984).
\bibitem{5} L. Arrachea, A.A. Aligia, E.R. Gagliano, K. Hallberg, and
C.A. Balseiro, Phys. Rev. B {\bf 50}, 16044 (1994), 
{\bf 52}, 9793 (1995) (E) 
\bibitem{clay} R. Torsten Clay, A.W. Sandvik, and D.K. Campbell, 
Phys. Rev. B {\bf 59}, 4665 (1999).
\bibitem{mizu} Y. Mizuno, T. Tohyama, S. Maekawa, T. Osafune, N. Motoyama,
H. Eisaki, and S. Uchida, Phys. Rev. B {\bf 57}, 5326 (1997).
\bibitem{fir}Yu. A. Firsov, V.N. Prigodin, and Chr. Seidel, 
Phys. Rep. {\bf 126}, 245 (1985).
\end{references}
\end{document}